# Reversible melting and equilibrium phase formation of (Bi,Pb)$_2$Sr$_2$Ca$_2$Cu$_3$O$_{10+d}$

E. Giannini [a], I. Savysyuk, V. Garnier, R. Passerini, P. Toulemonde and

R. Flükiger

DPMC - University of Geneva - 24, quai Ernest-Ansermet, CH-1211 Genève, Switzerland.

(a) corresponding author
   phone:    +41 22 7026578
   Fax:      +41 22 7026869
   E-mail:  enrico.giannini@physics.unige.ch

# ABSTRACT

The decomposition and the reformation of the $(Bi,Pb)_2Sr_2Ca_2Cu_3O_{10+d}$ ("Bi,Pb(2223)") phase have been investigated *in-situ* by means of High-Temperature Neutron Diffraction, both in sintered bulk samples and in Ag-sheathed monofilamentary tapes. Several decomposition experiments were performed at various temperatures and under various annealing atmospheres, under flowing gas as well as in sealed tubes, in order to study the appropriate conditions for Bi,Pb(2223) formation from the melt. The Bi,Pb(2223) phase was found to melt incongruently into $(Ca,Sr)_2CuO_3$, $(Sr,Ca)_{14}Cu_{24}O_{41}$ and a Pb,Bi-rich liquid phase. Phase reformation after melting was successfully obtained both in bulk samples and Ag-sheathed tapes. The possibility of crystallising the Bi,Pb(2223) phase from the melt was found to be extremely sensitive to the temperature and strongly dependent on the Pb losses. The study of the mass losses due to Pb evaporation was complemented by thermogravimetric analysis which proved that Pb losses are responsible for moving away from equilibrium and therefore hinder the reformation of the Bi,Pb(2223) phase from the melt. Thanks to the full pattern profile refinement, a quantitative phase analysis was carried out as a function of time and temperature and the role of the secondary phases was investigated. Lattice distortions and/or transitions were found to occur at high temperature in Bi,Pb(2223), Bi,Pb(2212), $(Ca,Sr)_2CuO_3$ and $(Sr,Ca)_{14}Cu_{24}O_{41}$, due to cation diffusion and stoichiometry changes.

The results indicate that it is possible to form the Bi,Pb(2223) phase from a liquid close to equilibrium conditions, like Bi(2212) and Bi(2201), and open new unexplored perspectives for high-quality Ag-sheathed Bi,Pb(2223) tape processing.

1. **INTRODUCTION**

Superconducting tapes and wires made with Ag-sheathed $(Bi,Pb)_2Sr_2Ca_2Cu_3O_{10+\delta}$ ("Bi,Pb(2223)") are, at present, the only high temperature superconductors in production for industrial applications. The high transition temperature ($T_c$~108-110 K) and the encouraging critical current densities carried by such conductors (>75 kA/cm$^2$ and ~40 kA/cm$^2$ in short and industrial lengths, respectively), render this the most suitable material among high temperature superconductors for large scale power applications [1,2]. Unfortunately, Ag-sheathed Bi,Pb(2223) conductors only approach, but not completely fulfil, the technical and economical requirements for widespread production and application [3]. The average critical current is still lower than the expected potential of the Bi,Pb(2223) compound, even if critical current can locally achieve values as high as 180 kA/cm$^2$ over 50-100 µm long regions [4]. Only a small portion of the ceramic core effectively carries the superconducting current and even the best samples carry quite a modest current because of voids, secondary phases, imperfect texture and poor grain connectivity [5,6].

For more than ten years, several research groups have investigated the extreme complexity of this compound and the material fabrication difficulties, in order to improve the performance of such conductors. This has lead to strong efforts and detailed investigations of the thermodynamics and the phase formation mechanism [7]. Even if some details of the Bi,Pb(2223) tape processing are still debated, one can state that the phase formation mechanism is understood. The Bi,Pb(2223) phase forms via a dissolution-recrystallisation process involving a liquid phase. Indications of the presence of a liquid phase assisting the Bi,Pb(2223) phase formation were revealed by the presence of amorphous regions after quenching [8-10]. Accurate SEM observations allowed Grivel et al. [11] to strongly support a nucleation-and-growth formation mechanism against a solid state intercalation mechanism [12]. The first direct proof of the presence of liquid occurring at high temperature as a consequence of the precursor decomposition was obtained by high temperature *in-situ* neutron powder diffraction (NPD) [13]. This technique measures the total amount of crystalline matter inside the silver sheath throughout the whole thermal treatment and thus follows its evolution as a function of temperature and time.

*In-situ* high temperature transmission diffraction experiments have recently proved to be very powerful techniques for monitoring the phase evolution and the texture development inside Ag-sheathed Bi,Pb(2223) tapes and have lead to interesting and unambiguous results (see [7] and [14] for a review). Frello et al. [15] studied in detail the diffraction peak profile from high-energy synchrotron X-ray diffraction and supported a liquid assisted nucleation-and-growth formation mechanism. Using the same experimental technique, Poulsen et al. [16] and Gottschalck-Andersen et al. [17] studied the Bi,Pb(2223) phase formation both in air and reduced oxygen partial pressure. High temperature *in-situ* neutron diffraction, which is more efficient than synchrotron radiation for a quantitative phase analysis, has been successfully used by Giannini et al. [13] and Fahr et al. [18] in air and reduced oxygen partial pressure, respectively.

Although it has been proved that the Bi,Pb(2223) phase forms from a liquid, this liquid is not stable and does not form homogeneously: its transient and local nature (strongly dependent on parameters which are not under control, like the local $p(O_2)$, density, liquid composition…) is responsible for the lack of structural homogeneity and for the limitation of the superconducting properties of the Ag-Bi,Pb(2223) tapes. The transient liquid assisting the Bi,Pb(2223) phase formation forms from the local decomposition of one or more secondary phases far from equilibrium [19]. The knowledge of the phase formation mechanism has lead to discover the intrinsic limits of the standard Powder-In-Tube processing route to fabricate Ag-sheathed Bi,Pb(2223) tapes.

In order to go beyond these limits and to improve Ag-sheathed Bi,Pb(2223) tapes, one should question the whole processing route and look at it from a different thermodynamic approach offering new and unexplored opportunities. Bi,Pb(2223) phase formation from a liquid being either stable or close to equilibrium conditions would open new perspectives for tape processing and industrial applications. Bi,Pb(2223) formation after partial melting in a strong magnetic field has been reported to improve density and grain alignment in bulk samples [20]. More recently, a partial reformation of Bi,Pb(2223) after melting was observed by Poulsen et al. [21] by means of *in-situ* synchrotron X-ray diffraction. Nevertheless, the thermodynamic conditions to form Bi,Pb(2223) on cooling from a melt, like Bi(2212) and Bi(2201), are still to be found. Only recently, researchers have succeeded in forming the first Pb-free Bi(2223) single crystals by a travelling floating zone method [22,23].

In this paper we report a series of high-temperature *in-situ* neutron powder diffraction (NPD) experiments performed in several runs over the last two years in order to study the decomposition and the formation of the Bi,Pb(2223) phase from the melt upon cooling, in sintered rods as well as in Ag-sheathed tapes. Investigations were carried out under various annealing atmospheres, in open furnaces as well as in sealed tubes. The reactions occurring at high temperature upon melting and cooling were directly observed and a quantitative phase analysis was carried out from a full pattern refinement of the powder diffraction patterns.

These measurements show that Bi,Pb(2223) can be reformed on cooling from the melt, provided that Pb-losses at high temperature are kept under control. The appropriate conditions for the Bi,Pb(2223) reformation have been found and are discussed in this paper. First attempts to apply a melting-retransformation treatment to the tape processing are discussed: the critical current can be partially recovered after melting and slow cooling.

2. **EXPERIMENTAL**

Ag-sheathed Bi,Pb(2223) tapes were made following the usual Oxide-Powder-In-Tube (OPIT) route, starting from a mixture of co-precipitated oxalates with a nominal composition $Bi_{1.72}Pb_{0.34}Sr_{1.83}Ca_{1.97}Cu_{3.13}O_{10+\delta}$ [24]. Due to the strong neutron absorption in silver and the low ceramic-to-silver ratio in multifilamentary tapes, monofilamentary tapes (with less Ag than multifilamentary ones) were used for this study. In addition, a particularly thin Ag tube was used in order to increase the diffraction contribution from the ceramic core. After the full reaction treatment, the tapes were ~3 mm wide and ~100 µm thick, with the thickness of the silver sheath being only 30-35 µm. Diffraction targets were made by cutting 1.5 m of tape, rolling it into coils of 5 mm in diameter and stacking several coils on top of each other to form a cylinder-like sample ~40 mm high. Such a target configuration allowed us to reduce the effects of the preferred orientation of the Bi,Pb(2223) platelets in the tape, thus improving the quality and the reliability of the pattern refinement.

Almost pure Bi,Pb(2223) powder with the overall composition $Bi_{1.84}Pb_{0.32}Sr_{1.84}Ca_{1.97}Cu_{3.00}O_{10+\delta}$, was obtained from commercial precursors after 5 days annealing at 857°C in air with intermediate grindings. The final powder, containing > 90% Bi,Pb(2223) and very low amounts of Bi(2201), Bi(2212) and (Ca,Sr)-cuprates (as revealed by X-Ray Powder Diffraction), was pressed in a Cold Isostatic Press at 10 kbar into rod shaped samples 40 mm high and 5 mm in diameter and encased in a 25 μm thick Ag foil.

Two different sample holders were used, depending on the desired atmosphere. For the experiments performed in an open system, the sample was hung inside a quartz chamber on the furnace axis, under a gas flow at controlled oxygen partial pressure, as shown in Fig. 1(a). This sample chamber guarantees the desired oxygen partial pressure at the sample position and the vacuum inside the furnace ($P = 10^{-4}$ Pa) needed to prevent the Ta heaters from oxidising. Two flowing gas mixtures were used: 99%Ar-1%$O_2$ and 80%Ar-20%$O_2$; the gas flow was fixed at 1.5 l/h. In a further experiment, samples were heat-treated inside sealed quartz tubes under an Ar-$O_2$ mixture at $p(Ar) = 0.8$ bar and $p(O_2) = 0.2$ bar at T = 850° C. Sealed quartz tubes were inserted into a cylindrical vanadium sample holder as shown in Fig. 1(b). The decision to close the samples inside quartz tubes was taken to reduce the Pb losses at high temperature, as discussed in Session 3.2.

Cylindrical targets (either rolled tapes or sintered rods) were treated in a dedicated furnace constructed with coaxial cylindrical tantalum foils, wrapped in an external aluminium cold screen. The diffraction contribution from the furnace was suppressed by a radial collimator up to a scattering angle $2\theta = 160°$. The furnace was equipped with two Pt/Pt-Rh10% thermocouples and the temperature variation over the region illuminated by the neutron beam was estimated to be $\Delta T < 2°C$.

NPD experiments were performed at the SINQ spallation source at the Paul Scherrer Institut (PSI) in Villigen (Switzerland), in successive runs between December 2000 and October 2001, using the High Resolution HRPT diffractometer. The optimum wavelength of the neutron beam was $\lambda=1.886$ Å for a take off angle of 120°. Neutrons were detected by a Position Sensitive Detector bank of 1600 proportional counters at intervals of 0.1° covering an angle of 160°. Thanks to the high scattering angles of the monocromators (120°) and of the sample (160°), the resolution is very high ($\delta d/d=0.001$) [25,26]. The scattering geometry is drawn in Fig. 2.

After a reference diffraction pattern was acquired at room temperature, the samples were first heated up to the temperature at which the Bi,Pb(2223) phase had been formed (825°C-855°C, depending on the sample), held at this temperature for 40-60 min for the acquisition of a high temperature reference diffraction pattern, and then heated up to the decomposition temperature. Different thermal profiles were used on cooling, depending on the kind of sample (bulk or tape) and on the oxygen partial pressure, and diffracted neutrons were continuously acquired. It is worthwhile to remember that in this kind of experiment, the chosen thermal profile is a compromise between the reactions occurring inside the sample and the general status of the experiment (beam intensity, acquisition rate, available beam time…), and it is therefore impossible to test a wide assortment of cooling procedures.

Diffraction patterns were analysed by a full pattern profile refinement technique based on the Rietveld method [27] using a DBWS program [28]. Pattern refinement and quantitative phase analysis were carried out as described in Ref. [13].

As an application of the NPD results, some Ag-sheathed tapes were subjected to a partial melting-reformation step and the effect on the critical current was monitored by standard four-probe I-V measurements.

Complementary DTA/TG measurements were performed in a commercial SETARAM TAG 24 equipment. XRD measurements were made in a PHILIPS PW-1820 powder diffractometer, using the standard Bragg-Brentano geometry.

## 3. RESULTS

### 3.1 Bi,Pb(2223) decomposition under gas flow

#### 3.1.1 Bulk samples

An example of acquired and calculated diffraction patterns is shown in Fig. 3. The diffraction pattern was acquired at T=850°C under $p(O_2)$=0.2 bar during ~40 min and the corresponding refinement $R_{WP}$ factor [28] was 7.7. Bi,Pb(2223) neutron reflections are indicated by vertical ticks at the bottom: at this stage the sample was almost single-phase Bi,Pb(2223). Starting from such diffraction patterns, by means of the full pattern profile refinement and following the same procedure as described in [13], we estimated the phase content as a function of temperature during melting and slow cooling. Very good agreement ($\Delta T \leq 1°C$) was found between the temperature at which the decomposition of the Bi,Pb(2223) phase was observed by means of diffraction experiments and the decomposition temperature measured by Differential Thermal Analysis (DTA) on a sample of the same composition; this indicates the reliability of the temperature control in our experiments.

The gradual decomposition of Bi,Pb(2223) in a sintered rod heated at 880°C under a flowing 20%$O_2$-80%Ar atmosphere is shown in Fig. 4(a). The main secondary phase forming at the Bi,Pb(2223) decomposition was found to be $(Ca,Sr)_2CuO_3$ and it is plotted in Fig. 4(b).

The absolute amount of the *j*-th phase is calculated by multiplying the refined scale factor (*S*) by the number of formula units per unit cell (*Z*), the mass of the formula unit (*M*) and the volume of the unit cell calculated from the refined cell parameters (*V*) (see [13]):

$$w_j = S_j Z_j M_j V_j \qquad (1)$$

This quantity does not depend on any normalisation and directly expresses the amount of each crystalline phase present in the sample at a given temperature and time. The temperature vs. time profile is superposed on each graph. Once a temperature of 880°C is reached (a temperature chosen because it is above the melting point under these conditions), the sample was step-cooled down to 875°C, 873°C, 872°C and then cooled at a constant rate, $dT/dt = 3$°C/h, down to 850°C (below this temperature the sample was fast cooled at 600°C/h down to room temperature). The choice of such a thermal profile was found to be disadvantageous, as the Bi,Pb(2223) phase decomposed too rapidly, and was therefore not repeated for the other samples. At high temperature (T > 875°C), only Bi,Pb(2223) and $(Ca,Sr)_2CuO_3$ contribute appreciably to the diffraction pattern, whereas the other secondary phases (namely $(Sr,Ca)CuO_2$, Bi(2201) and Bi(2212)) are present in trace amount, and are thus difficult to quantify. As is clearly visible in Fig. 4, the Bi,Pb(2223) phase did not form from the melt on cooling under the chosen conditions.

The experiment was repeated in reduced oxygen partial pressure under a flowing 1%$O_2$-99%Ar atmosphere. The Bi,Pb(2223) and $(Ca,Sr)_2CuO_3$ phase contents are shown in Fig. 5(a) and 5(b), respectively. The heating profile, the decomposition temperature and the dwell time at high temperature were chosen to achieve a partial decomposition of Bi,Pb(2223). However, no re-crystallisation of Bi,Pb(2223) was observed upon cooling and the trend of the main secondary phase, $(Ca,Sr)_2CuO_3$, was found to be the same as in Fig. 4(b).

A partial decomposition of Bi,Pb(2223) occurs on heating before the melting temperature is reached, depending on the heating rate. According to the phase diagram proposed by Strobel et al. [29], between 860°C and 876°C the Bi,Pb(2223) phase should coexist with the Bi(2212) phase and some liquid phase before melting incongruently into Bi(2201), cuprates and liquid phase. As pointed out by Sastry et al. [30] and confirmed by Park et al. [31], the Bi,Pb(2223) phase is a solid solution, and therefore the melting temperature covers quite a wide range instead of occurring at a sharp point. Garnier et al. [32] have recently discussed the evolution with temperature of the equilibrium Bi,Pb(2223)-Bi(2212)-Bi(2201) and have confirmed a decrease of the Bi,Pb(2223) content above 860°C, well before the incongruent melting of this phase takes place. The behaviour of the Bi(2212) phase therefore differs from sample to sample, depending on the Bi,Pb(2223)-Bi(2212) equilibrium. In the sample melted under $p(O_2) = 0.2$ bar (Fig. 4), Bi(2212) was already formed on heating between 860°C and 880°C, but its state did not change as a result of the Bi,Pb(2223) melting, and the Bi(2212) amount only started to increase after cooling below 850°C. In the sample melted under $p(O_2)$=0.01 bar (Fig. 5), no Bi(2212) phase was formed on heating. However, it precipitated on cooling below 830°C as shown in Fig. 6.

Other secondary phases could not be quantified because of the low signal-to-noise ratio at high temperature after the Bi,Pb(2223) decomposition. Reflection peaks of minority phases were often hidden in the background noise, while a large amount of the sample was amorphous or in a liquid state. In addition, the acquisition time was rather short (~30 min) due to the large temperature variations occurring at the same time. As a consequence, the refinement $R_{WP}$ factor was quite high in this first series of experiments ($R_{WP} \approx 15$). The neutron statistics will be improved in the second series of experiments, as discussed in Section 3.2.

### 3.1.2 Ag-sheathed tapes

In the case of Ag-sheathed monofilamentary tapes, the intensity of the diffraction peaks was lower than for bulk samples, the effective target volume being smaller and the diffraction patterns being dominated by the strong reflections of Ag. The acquisition of each pattern was extended over a longer time period, resulting in a broadening of the curve of phase contents as a function of temperature. The results obtained on monofilamentary tapes in 20%$O_2$-80%Ar are shown in Fig. 7. In Ag-sheathed tapes, the Bi,Pb(2223) phase is found to form directly from the melt during cooling : after quenching from 860°C to 850°C, the Bi,Pb(2223) content starts to increase and goes on increasing while cooling at a rate of 6°C/h (Fig. 7(a)). In the same temperature range $(Ca,Sr)_2CuO_3$, which remains the main secondary phase, partially decomposes (Fig. 7(b)). It should be noted that the phase contents obtained in our experiment are absolute amounts and not relative fractions: the decrease of $(Ca,Sr)_2CuO_3$ shown in Fig. 7(b) is thus real and not due to any artifice of normalisation.

The other secondary phases found in tapes are not the same as those found in sintered pellets : instead of traces of $(Ca,Sr)CuO_2$, we found a significant amount of $(Sr,Ca)_{14}Cu_{24}O_{41}$ (up to 15% weight fraction). This phase is known to be the main secondary phase in equilibrium with Bi,Pb(2223) in Ag-sheathed tapes [13] and was already present inside the tapes at the beginning of the experiment. Its amount ranges from 0.05 to 0.15 weight fraction and, even if its trend as a function of time and temperature is not as clear as for $(Ca,Sr)_2CuO_3$, it is likely to participate in the melting-reformation reaction.

Because we can directly measure the absolute amount of each phase, we could independently follow the behaviour of each of them and can infer that, under these experimental conditions and to a partial extent, Bi,Pb(2223) can be melted and formed reversibly from the melt following the reaction:

$$\text{Bi,Pb(2223)} \rightleftarrows (Ca,Sr)_2CuO_3 + (Sr,Ca)_{14}Cu_{24}O_{41} + liquid \qquad (2)$$

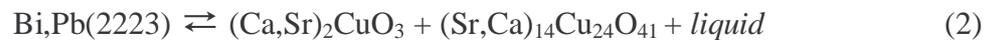

On the base of the *in-situ* diffraction results, monofilamentary Ag-sheathed tapes were subjected to partial melting at 860°C and a slow cooling in air in order to see the effect on the critical current. Various melting times and cooling rates were used. The starting critical current of $I_c = 13$ A ($J_c = 21 \cdot 10^3$ A/cm$^2$) dropped to 1.6 A after 1h at 860°C, but more than 50 % of the starting critical current value was recovered on cooling following the thermal profile in Fig. 7 ($I_{final} = 6.7$ A).

### 3.1.3 Effect of Pb losses.

The obvious question arising from the results presented so far is why does the Bi,Pb(2223) phase partially re-form after melting inside Ag-sheathed tapes but not in sintered bulk samples? Some differences between bulk and tapes can be discussed, taking into account the presence of Ag and the different density of the tapes.

All the measured samples exhibited a mass loss (up to 2%) at the end of the thermal treatment. Such mass losses are commonly attributed to the evaporation of Pb, the most volatile cation in a Bi,Pb(2223) sample [33]. In order to investigate the mass losses occurring at the phase decomposition, samples from the same batches as used for diffraction experiments were annealed in a DTA/TG analyser under the same conditions as used for NPD, and their mass was measured during the whole thermal treatment. The same thermogravimetric measurements were repeated on pure Au reference samples in order to take into account any weight changes due to pressure and crucible effects. The results of the sintered rod melted in 20%$O_2$-80%Ar are shown in Fig. 8. Several steps of mass loss take place before melting, with most of the loss occurring at the decomposition temperature, where the sample looses 1% of its mass in 1 h. A slow, steady, mass increase was measured on cooling between 870°C and 850°C, which we attribute to the inlet of the oxygen needed for precipitation of secondary phases like Bi(2212) and Bi(2201). At the end of the cooling the mass loss was only partially recovered, showing that some irreversible loss has occurred. Mass losses in sintered rod melted under reduced oxygen partial pressure ($p(O_2) = 0.01$ bar) exhibited the same trend but they were more pronounced.

In Ag-sheathed tapes the mass losses are appreciably lower. In Fig. 9 we compare the mass losses occurring in a sintered rod and in a tape upon heating in 20%$O_2$-80%Ar, before melting at 880°C and 860°C, respectively. Losses are normalised on the same scale, taking into account the effective mass of the ceramic core of the tape. One can see that in Ag-sheathed tapes the mass loss is reduced by almost a factor 2. At the end of the experiment, SEM/EDX analysis confirmed a lower Pb content in the bulk samples than in the tapes (~2 and ~3 atomic % respectively).

We attribute the irreversibility of melting to a change in the stoichiometry due to Pb losses, and hence to a movement from one multi-phase region of the phase diagram to another. In Ag-sheathed tapes, the density of the ceramic core was higher than in sintered rods: liquid phase was prevented from flowing away and a more homogenous composition was kept throughout the sample during the melting process. Moreover, the Ag sheath can play an important role in preventing the Pb evaporation and homogenising the liquid-solid interface, thus favouring the Bi,Pb(2223) reformation on cooling.

Similar conclusions had been reached in our previous study of Bi,Pb(2223) stability under high isostatic pressure [33]. An indirect confirmation of the negative role of Pb losses on Bi,Pb(2223) formation from the melt comes from crystal growth studies. At present, the only large and high-quality Bi(2223) single crystals have been grown from a Pb free precursor [22,23].

### 3.2    Bi,Pb(2223) decomposition in sealed tubes.

As reported in [33], an efficient way to reduce Pb losses and to control the stoichiometry during melting and cooling is to work under isostatic pressures as high as 10-30 MPa. Building a high-pressure neutron compatible furnace working under oxidising atmosphere is not a trivial technical challenge and so we opted for enclosing samples inside sealed quartz tubes as shown in Fig. 1(b).

Almost single-phase Bi,Pb(2223) rods, as used for the previous experiment, were enclosed inside quartz tubes under controlled low pressure atmosphere, in order to set $p(O_2)$=0.2 bar and $p(Ar)$=0.8 bar at 850°C. The volume of the quartz sample holder was as small as possible to minimise the gas volume surrounding the sample and to saturate the atmosphere with Pb. Diffraction experiments were performed as described above, but the thermal profiles were modified, replacing slow temperature ramps with dwell times at constant temperature. The acquisition time could be prolonged and the quality of the diffraction patterns were noticeably improved. Furthermore, during this second series of experiments, the neutron beam was more intense than before, allowing better statistics and thus improving the general quality of the NPD patterns. Refinement $R_{WP}$ factors were found to be quite low, ranging from 5 (for 90 min acquisition below the melting temperature) to 9.4 (for 20 min acquisition at 894°C).

Various annealing temperatures, heating and cooling rates and dwell times at a given temperature were tested and the reformation of Bi,Pb(2223) upon cooling was observed. Results from two different experiments starting from the same precursor are shown in Fig. 10 and 11. Fig. 10(a) shows an example of a sample where the Bi,Pb(2223) phase did not form from the melt on cooling, but Fig. 11(a) shows a successful experiment where Bi,Pb(2223) did reform.

In these two runs, we carried out several subsequent heating steps above the melting point and cooled down to various annealing temperatures, with the aim of looking for the equilibrium temperature range between the solid and the liquid phase, i.e. a solidus line in the phase diagram. For the sake of clarity, only two of these high-temperature cycles are shown in Fig. 10 and 11. The sample in Fig. 10 was heated from 850°C to 894°C in 3 h, then quenched down to 857°C and annealed for ~15h at this temperature. After two steps (1 h) at 862°C and 867°C, the samples was fast heated to 894°C and quenched down to 845°C for a second annealing. For the thermal profile in Fig. 11, different heating rates and annealing temperatures were chosen. After a first fast heating to 894°C and cooling (in less than 2 h), the sample was annealed for ~6 h at 862°C, then heated up to 894°C-896°C at 10°C/min, held for 2 h above the melting point and quenched down to 862°C. During the second annealing at 862°C (~6 h) the Bi,Pb(2223) phase was reformed. During further heating and annealing steps at the same temperature (not shown in Fig. 11), the decomposition and retransformation were found to occur reversibly and the result was reproduced. Annealing steps at higher temperatures (867°C and 870°C) did not allow the Bi,Pb(2223) phase to reform, showing that the appropriate temperature range for the Bi,Pb(2223) phase formation is very narrow.

Fig. 10 (b) and (c), and Fig. 11(b) and (c) show the behaviour of the secondary phases Bi,Pb(2212), $(Sr,Ca)_{14}Cu_{24}O_{41}$ and $(Ca,Sr)_2CuO_3$. Bi(2201) was only observed in some patterns and its behaviour will be discussed later. Other secondary phases like $(Ca,Sr)CuO_2$, observed in samples melted in open systems (see Section 3.1), did not crystallise in any sample treated in sealed tubes.

### 3.2.1 Phase analysis during melting and cooling

We now discuss in detail the phase evolution as a function of time and temperature. After the first decomposition stage in Fig. 10, the sample is cooled down to 855°-857°C (the temperature at which Bi,Pb(2223) was previously synthesised) as fast as possible and then kept at constant temperature for several hours. The decomposition of Bi,Pb(2223) occurs quite rapidly during heating and no phase reformation is observed at constant temperature. During the same time, the very small amount of Bi,Pb(2212) present at the beginning decomposes, but after cooling the phase re-crystallises in larger amounts (up to 0.50 weight fraction), as shown in Fig. 10(b). Secondary cuprates start to grow as soon as the temperature is increased above 850°C (Fig. 10(c)). At 894°C the weight fractions of crystalline $(Sr,Ca)_{14}Cu_{24}O_{41}$ and $(Ca,Sr)_2CuO_3$ are 27% and 16% respectively. Both of them decompose only partially on cooling, and settle at values of 16% and 11% respectively during the annealing step at constant temperature. No appreciable phase evolution is measured after ten hours at 857°C.

The rod is then subjected to a second short cycle at 894°C before being cooled down to 845°C. During this second melting-cooling run, the residual Bi,Pb(2223) decomposes almost completely and will never reform (see Fig. 10(a)). Bi,Pb(2212) also decomposes partially above 867°C, but re-crystallises after cooling to 845°C, becoming the majority phase at the end of the treatment (Fig 10(b)). The behaviour of the two cuprates $(Sr,Ca)_{14}Cu_{24}O_{41}$ and $(Ca,Sr)_2CuO_3$ is similar to that at the first heating at 894°C, as shown in Fig. 10(c).

We deduce from these results that the temperatures chosen for the annealing after melting and cooling were too low, and that the first heating above the melting point was too slow (strong decomposition of Bi,Pb(2223) and formation of cuprates took place before the maximum temperature was reached). As a consequence, the phase equilibrium moves toward a stable phase composition of Bi,Pb(2212) + $(Sr,Ca)_{14}Cu_{24}O_{41}$ + $(Ca,Sr)_2CuO_3$ and no Bi,Pb(2223) can be recovered.

In the experiment illustrated in Fig. 11, the temperature was rapidly increased and the annealing temperature after cooling was higher. For the sake of clarity of the presentation, the vertical axis scale has been changed. In Fig. 11(a), the initial amount of Bi,Pb(2223) is out of scale: the initial decomposition occurring during heating is not shown. Less than 50% of the initial amount of Bi,Pb(2223) has decomposed before cooling down to 862°C. A correspondingly lower amount of secondary phases have formed (Fig. 11 (b) and (c)), even if the trend of $(Sr,Ca)_{14}Cu_{24}O_{41}$ and $(Ca,Sr)_2CuO_3$ is clearly the same as shown in Fig. 10(c). During the first step at 862°C no phase evolution is visible and Bi,Pb(2223) is in a multi-phase equilibrium with Bi,Pb(2212) and cuprates. The most interesting feature is the decomposition and reformation occurring during the second heating and the subsequent annealing at 862°C. After a partial melting of Bi,Pb(2223) corresponding to one half of the phase content, the superconducting phase is recovered and correspondingly the secondary phases formed at high temperature decompose. After 7 hours at 862°C on the second annealing step, the phase composition of the sample is almost the same as before melting. This proves that at such a temperature, under these experimental conditions, the sample is in an equilibrium state. Crossing a solidus to move in a multiphase pocket of the phase diagram containing $(Sr,Ca)_{14}Cu_{24}O_{41}$, $(Ca,Sr)_2CuO_3$ and a liquid turned out to be a reversible phase transition. Under these conditions, we obtained in bulk samples the same result as previously obtained in Ag-sheathed tapes.

The Bi,Pb(2223) phase reformation seems to occur following two different regimes, marked by two slopes of the phase amount vs. time (Fig. 11(a)). By using the volume fraction of the growing Bi,Pb(2223) phase we can use Avrami's equation to extract information about the crystallisation mechanism [34-36]. The first stage of phase reformation is found to correspond to an Avrami coefficient of $n \approx 3$, a very high value which is associated with a 3-D nucleation from a liquid at a constant rate. The second, and slower, stage of crystallisation is found to correspond to $n \approx 0.5$, which indicates a growth and a thickening of 2-D platelets. The same value is commonly reported for the late stage of Bi,Pb(2223) grain growth starting from the usual precursors [15,36,37]. The details of the kinetic studies of the reformation will be reported elsewhere.

The last point to discuss is the presence and the role of Bi(2201). Quantitative analysis of this phase turned out to be very difficult and results are sometimes contradictory. The Bi(2201) phase was present in very low amounts, generally one order of magnitude lower than that of Bi(2212). At such low amounts the quantitative phase analysis is not very reliable. However, we can say that, contrary to the Bi,Pb(2212) phase, the Bi(2201) content was found to increase at the Bi,Pb(2223) decomposition temperature and to decrease during Bi,Pb(2223) reformation at lower temperature. This is in agreement with the phase diagram proposed by other authors [29,31,38].

### 3.2.2 Structural studies

The good quality of data obtained in the diffraction experiments performed in sealed tubes allows us to further the study of the reactions occurring at high temperature, by performing lattice parameter and texture refinement. By plotting the refined lattice parameters as a function of time, some structural transitions come to light.

In Fig. 12, $a = b$ and $c$ lattice parameters of Bi,Pb(2223) are plotted vs. time over the same time scale as in Fig. 11. Corresponding to the Bi,Pb(2223) phase reformation occurring on the second step at 862°C, an increase of the unit cell volume is observed ($\Delta a = + 0.002$ Å, $\Delta c = + 0.02$ Å), indicating an irreversible stoichiometric change occurring in the crystallising phase. Such a phenomenon is not observed in a sample where the Bi,Pb(2223) phase does not reform on cooling.

The behaviour of Bi,Pb(2212) is clearly different, as shown in Fig. 13. The weak orthorhombic distortion, indicating a very small Pb content, is enlarged during the melting of the Bi,Pb(2223) phase but is reversible on cooling back to 862°C. The same behaviour is observed in other samples.

Finally, the secondary cuprates, also exhibit some lattice parameter changes, probably due to a variation in the Ca:Sr ratio at high temperature. A general increase of the *c* parameter and a decrease of the *b* parameter of $(Ca,Sr)_2CuO_3$ have been measured, as well as some irreversibility in the *a* and *b* parameter state of $(Sr,Ca)_{14}Cu_{24}O_{41}$. Actually, wide solid solutions of these two phases exist, and Sr and Ca diffusion at high temperature is likely to be possible (see [39-41] for phase diagram studies of these compounds).

## 4. DISCUSSION

We now summarise the results of the present study. The main result is that it is possible to form the Bi,Pb(2223) phase directly from the melt and that the phase decomposition is reversible under appropriate thermodynamic conditions. We have obtained the reformation of the Bi,Pb(2223) phase both in Ag-sheathed tapes and in sintered rods, but under different conditions. In order to make the decomposition reversible in bulk samples, it has been necessary to work in sealed tubes and to accurately control the heating rate and the annealing temperature. In Ag-sheathed tapes the Bi,Pb(2223) phase reformation was observed in an open system, but to a lesser extent. In all cases, small mass losses have been measured in the samples where the Bi,Pb(2223) phase reformation took place. By minimising the evaporation of volatile elements (we assume that the only volatile element is Pb), the stoichiometry of the sample is kept closer to the initial value and this constitutes the base condition for reversibility. Some Pb evaporation takes place in sealed tubes too, and some PbO has been found on the inner wall of the quartz tube, but the losses are minimised by saturating the atmosphere with Pb. The total mass loss measured in the samples treated in sealed tubes was only ~0.15%, even after four excursions above the melting point and a total annealing time of 36 hours.

In the samples exhibiting large mass losses and no Bi,Pb(2223) reformation, fast growth of secondary cuprates, mainly $(Ca,Sr)_2CuO_3$, has been observed and the multiphase equilibrium moves from a composition Bi,Pb(2223) + $(Ca,Sr)_2CuO_3$ + $(Sr,Ca)_{14}Cu_{24}O_{41}$ + *liquid* to a composition Bi,Pb(2212) + $(Ca,Sr)_2CuO_3$ + *liquid*. We are dealing with a very complex multiphase *solid-liquid-vapour* equilibrium covering a narrow range of the phase diagram.

We note that, even using sealed tubes and keeping the stoichiometry close to its starting value, the Bi,Pb(2223) phase does not reform from the melt at the same temperature used to synthesise the phase from the usual precursors, T=857°C (Fig. 10). The reaction temperature increases, and the Bi,Pb(2223) phase forms from the melt at T=862°C. This is in agreement with our predictions based on high pressure equilibrium studies on the Bi-Pb-Sr-Ca-Cu-O system, previously reported in [33] : it was found that under high isostatic pressure and lower Pb losses, the stability edge of Bi,Pb(2223) shifts to higher temperatures. The reformation range was found to be very narrow: no Bi,Pb(2223) crystallisation was observed after heating above 867°C.

Finally, we discuss our results in comparison with the earlier work of Park et al. [31] performed by *in-situ* high-temperature X-ray diffraction on Pb-free Bi(2223). We found the same peritectic reaction involving Bi,Pb(2223), cuprates and a liquid phase, in bulk samples as well as in Ag-sheathed tapes, i.e. both with and without Ag. Furthermore, using neutron diffraction, we were able to identify and quantify the secondary phases and to observe lattice parameters changes occurring at high temperature which prove that a mutual exchange of cations among different solid solutions takes place.

## 5. CONCLUSIONS

In this paper we have reported several *in-situ* high temperature neutron powder diffraction experiments performed on Bi,Pb(2223) bulk and tapes in order to study the phase decomposition and the Bi,Pb(2223) phase reformation from the melt. Bi,Pb(2223) was found to melt incongruently into $(Ca,Sr)_2CuO_3$, $(Sr,Ca)_{14}Cu_{24}O_{41}$ and a Bi,Pb-rich liquid. Pb losses were found to play a key role for the phase equilibria and the reversibility of the Bi,Pb(2223) decomposition process. In order to prevent Pb from evaporating and to form the Bi,Pb(2223) phase directly from the melt, thermal treatments in sealed quartz tubes were successfully used. The melting-reformation process can be represented, both with and without Ag, by the following reaction :

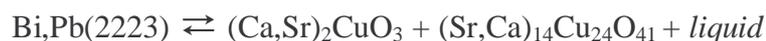

Bi,Pb(2223) $\rightleftarrows$ $(Ca,Sr)_2CuO_3$ + $(Sr,Ca)_{14}Cu_{24}O_{41}$ + *liquid*

Bi,Pb(2212) was not found to take part in this equilibrium process. This is in contrast with the non-equilibrium Bi,Pb(2223) phase formation in the OPIT process, where Bi,Pb(2212) plays a dominant role. On the other hand, Bi(2201) is involved, but in very low amounts and its role has still to be elucidated. The appropriate conditions for Bi,Pb(2223) reformation are extremely sensitive to the temperature and the local Pb content.

Interesting lattice distortions were observed at high temperature during the melting-reformation process in Bi,Pb(2223), $(Ca,Sr)_2CuO_3$ and $(Sr,Ca)_{14}Cu_{24}O_{41}$. Our data revealed stoichiometry changes in the solid solutions following the formation and the crystallisation of the melt.

Transport measurements performed on monofilamentary Ag-Bi,Pb(2223) tapes after partial decomposition and reformation show that the grain links and the critical current are partially recovered.

In conclusion, we have found that the equilibrium formation of Bi,Pb(2223) can be controlled. The existence of a new equilibrium reaction path for Bi,Pb(2223) phase processing has been demonstrated. The next step is to apply these results to Ag-Bi,Pb(2223) tapes, with the hope of homogenising the final filaments and thus improving the transport properties for large scale applications.

**ACKNOWLEDGEMENTS**

The authors gratefully acknowledge Dr. P. Fischer and Dr. D. Sheptyakov of the Laboratory for Neutron Scattering ETHZ-PSI Villigen (CH) for their help during the data acquisition, and Dr. M. Zolliker (LNS/PSI) for the help with the high-temperature environment. The authors wish to thank Dr M. Lomello-Tafin for useful discussions and Dr. N. Clayton for carefully reading the manuscript.

This work was supported by the Swiss National Research Fund and the National Centre of Competence in Research (NCCR) on Materials with Novel Electronic Properties (MaNEP).

**FIGURE CAPTIONS**

Figure 1.

(a) Quartz sample holder used for NPD experiments under a flowing $O_2$-Ar gas mixture. The gas flow inside the sample holder maintains the required $p(O_2)$ at the sample position. The thermocouple is placed in contact with the upper side of the sample, inside the inner quartz rod.

(b) Vanadium sample holder used for NPD experiments with sealed tubes. The oxygen partial pressure inside the quartz tube was set to be $p(O_2) = 0.2$ bar at $T = 850°C$. The temperature is measured at two different positions $T_1$ and $T_2$.

Figure 2.

Scattering geometry of the HRPT diffractometer. The high take-off angle (120°) and the wide angle range of the detector (160°) improve the resolution.

Figure 3.

NPD pattern of a bulk Bi,Pb(2223) sample at 830°C. The refined pattern (solid line) is superposed on the acquired data (symbols). On the bottom, reflections of Bi,Pb(2223) and a difference curve are shown.

Figure 4.

Bi,Pb(2223) *(a)* and $(Ca,Sr)_2CuO_3$ *(b)* phase content as a function of time during cooling after melting at 880°C in 20%$O_2$-80%Ar.

The temperature vs. time profile (solid line) is added. Dashed lines are guidelines.

Figure 5.

Bi,Pb(2223) *(a)* and $(Ca,Sr)_2CuO_3$ *(b)* phase content as a function of time during cooling after melting at 880°C in 1%$O_2$-99%Ar. The temperature vs. time profile (solid line) is added. Dashed lines are guidelines.

Figure 6.

Bi,Pb(2212) phase content as a function of time during cooling after melting at 860°C in 1%$O_2$-99%Ar. The phase does not form at the Bi,Pb(2223) decomposition, but at a lower temperature (T < 830°C) during cooling. The temperature vs. time profile (solid line) is added. Dashed lines are guidelines.

Figure 7.

Bi,Pb(2223) *(a)* and $(Ca,Sr)_2CuO_3$ *(b)* phase content inside an Ag-sheathed tape during cooling after melting at 860°C in 20%$O_2$-80%Ar. *In-situ* reformation of Bi,Pb(2223) is observed, along with a corresponding decomposition of $(Ca,Sr)_2CuO_3$. The temperature vs. time profile (solid line) is added. Dashed lines are guidelines.

Figure 8.

Thermogravimetric analysis of a Bi,Pb(2223) sintered pellet melted at 880°C under a flowing 20%$O_2$-80%Ar atmosphere. Strong mass losses occur at high temperature and are only partially recovered on cooling.

Figure 9.

Comparison of the mass losses occurring in Bi,Pb(2223) sintered pellets (filled symbols) and Ag-sheathed tapes (hollow symbols). The temperature vs. time profiles are added: sintered pellet = solid line and tape = dashed line. The differences in the observed mass losses are ascribed to a difference in Pb-evaporation in bulk samples and Ag-sheathed tapes.

Figure 10.

Bi,Pb(2223) *(a)*, Bi,Pb(2212) *(b)* and (Ca,Sr)-cuprate *(c)* phase content in a sintered bulk sample treated in sealed quartz tubes in a 20%$O_2$-80%Ar atmosphere. At T = 855°C after cooling, no Bi,Pb(2223) reformation from the melt is observed and a large amount of secondary phases are formed. Temperature vs. time profile (solid line) is added.

Figure 11.

Bi,Pb(2223) *(a)*, Bi,Pb(2212) *(b)* and (Ca,Sr)-cuprate *(c)* phase content in a sintered bulk sample treated in sealed quartz tubes in a 20%$O_2$-80%Ar atmosphere. At T = 862°C after cooling, Bi,Pb(2223) reformation from the melt is observed. A corresponding decomposition of (Sr,Ca)-cuprates occurs. Temperature vs. time profile (solid line) is added.

Figure 12.

Refined *a* = *b* and *c* lattice parameters of Bi,Pb(2223) as a function of time during melting and recrystallisation. Temperature vs. time profile (solid line) is added.

Figure 13.

Refined *a* and *b* lattice parameters of Bi,Pb(2212) as a function of time. A strong orthorhombic distortion appears at the Bi,Pb(2223) decomposition. Temperature vs. time profile (solid line) is added.

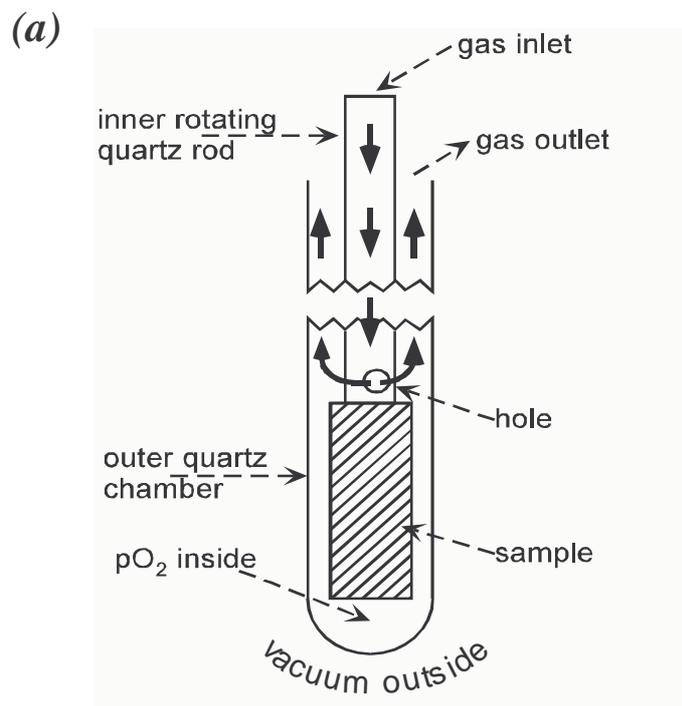 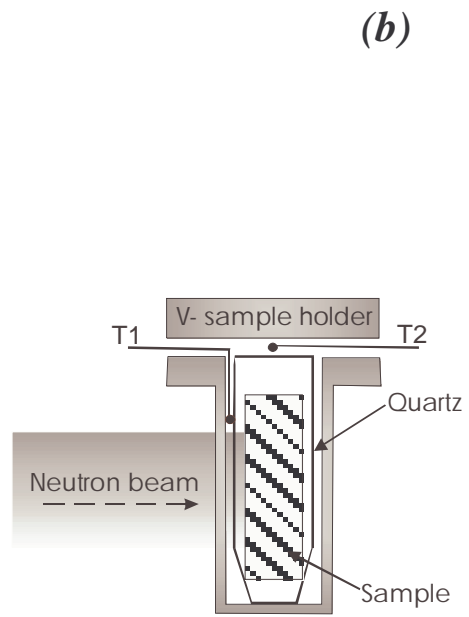

Fig. 1

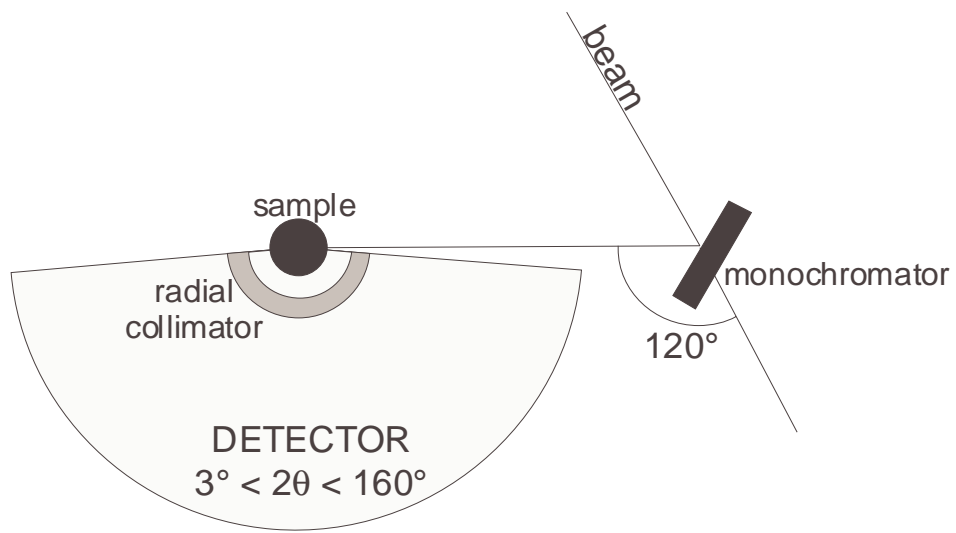

Fig. 2

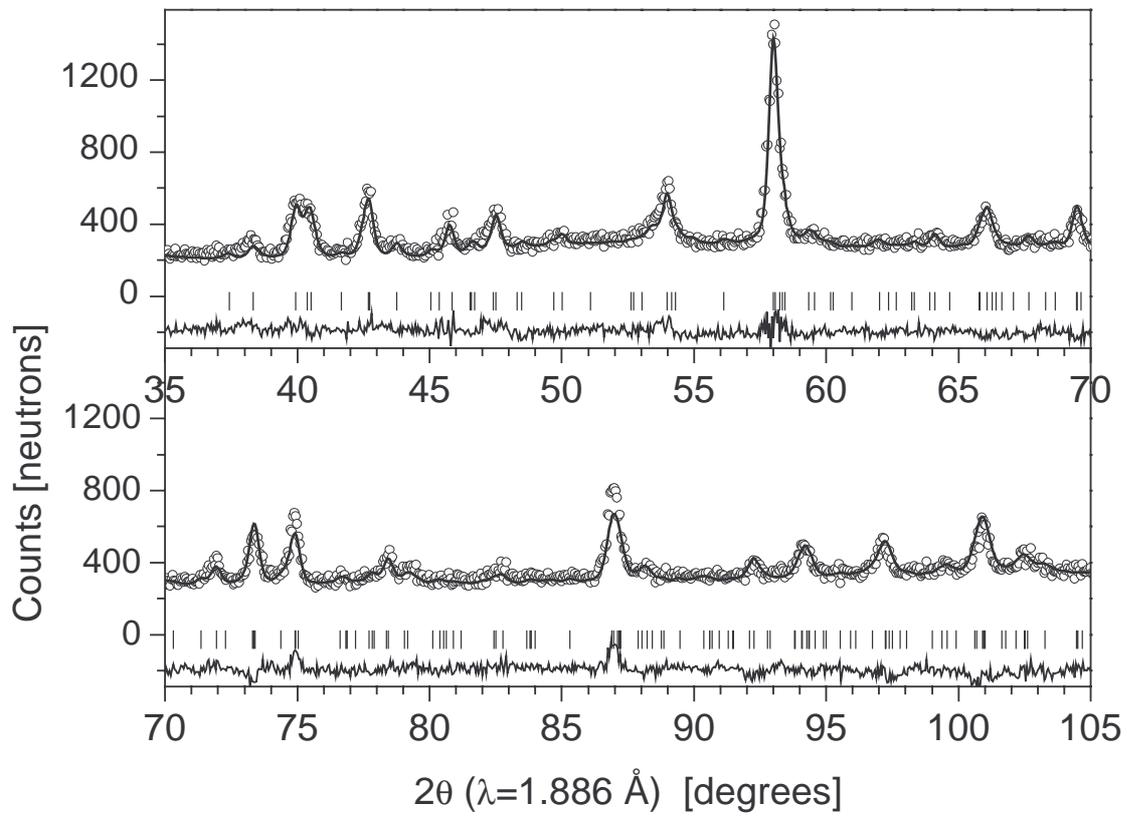

Fig. 3

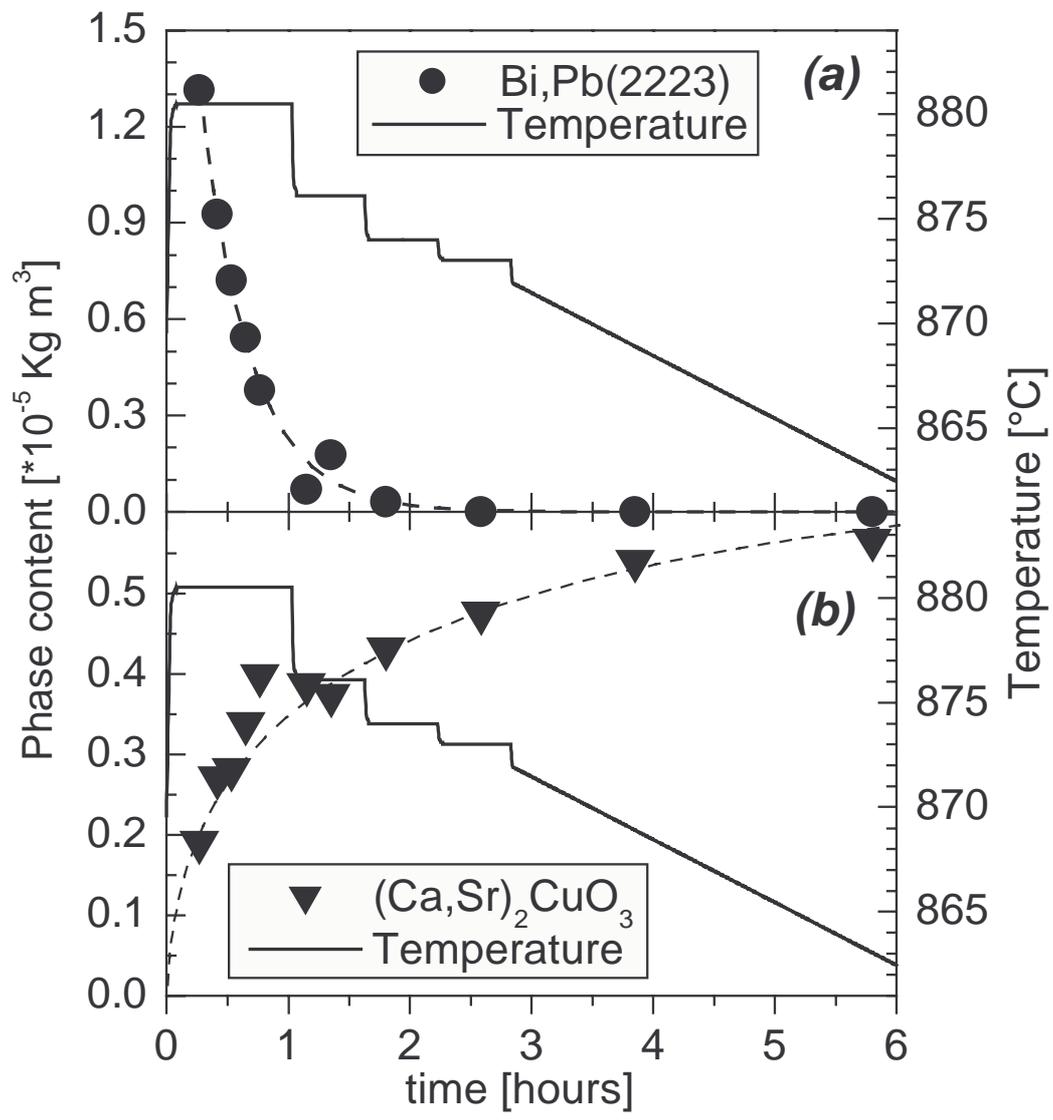

Fig. 4

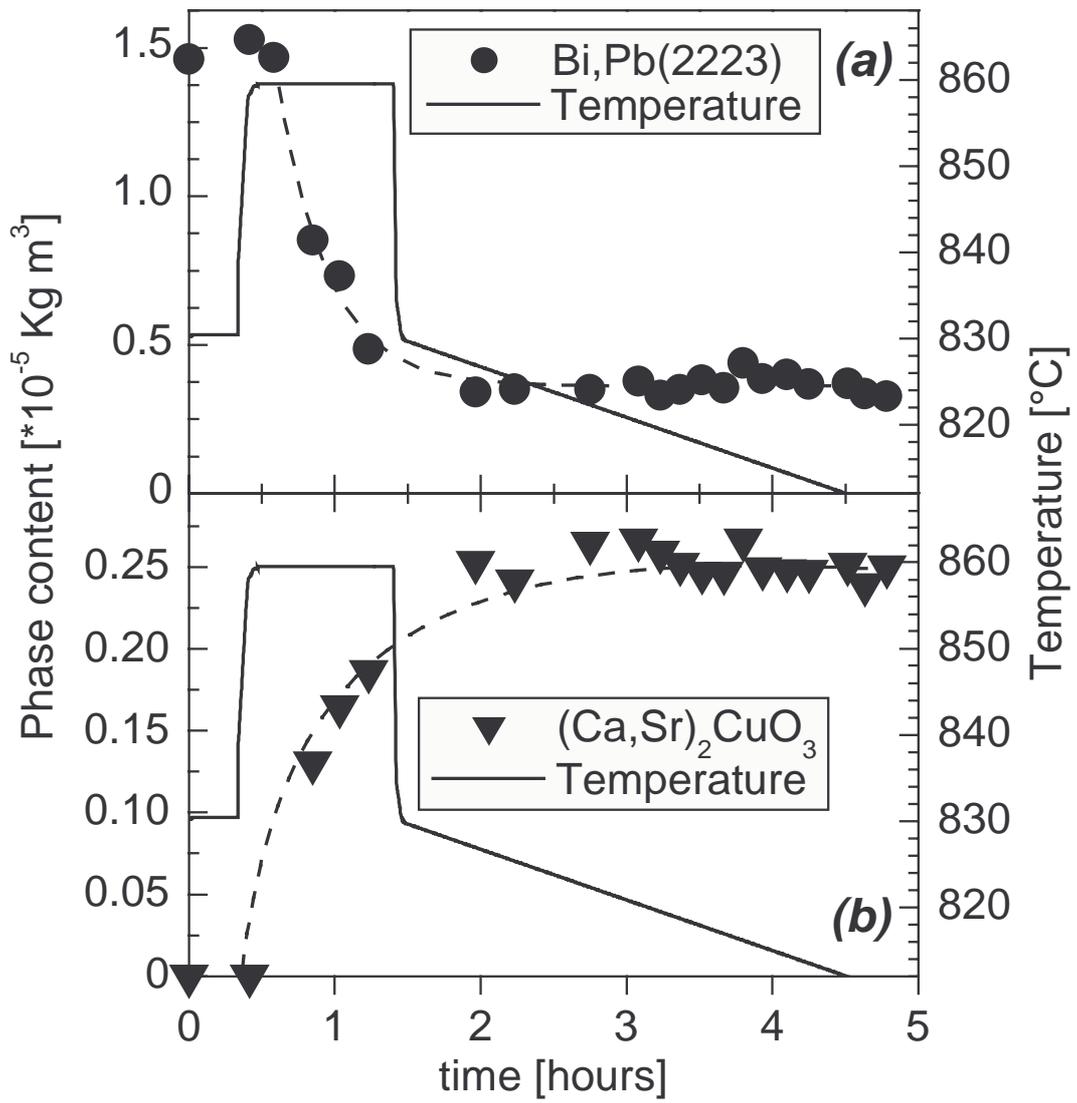

Fig. 5

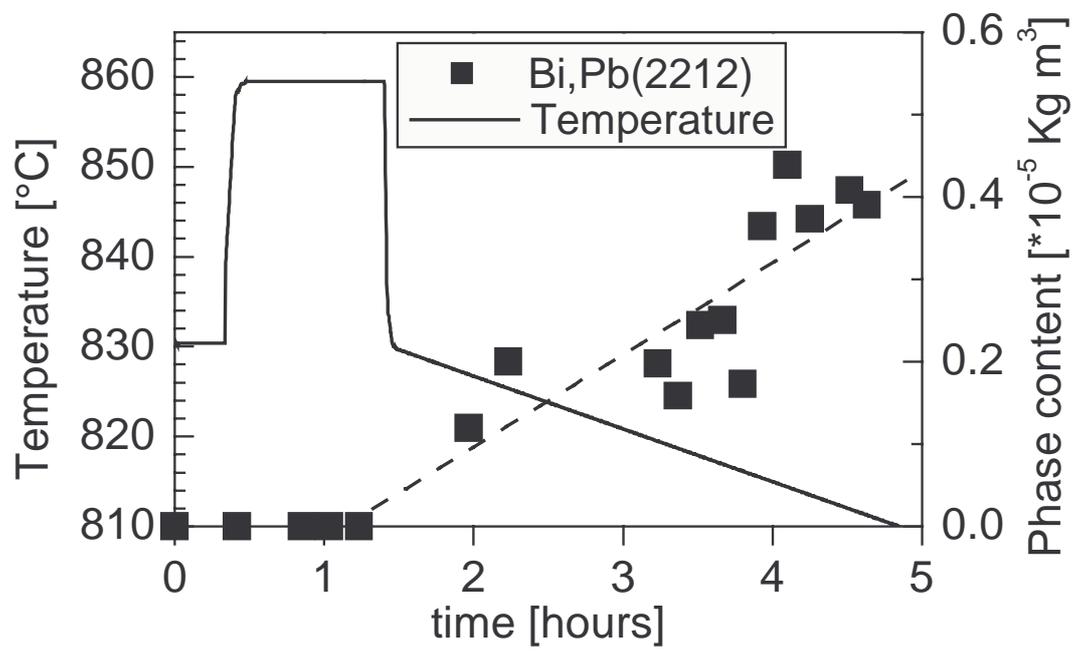

Fig. 6

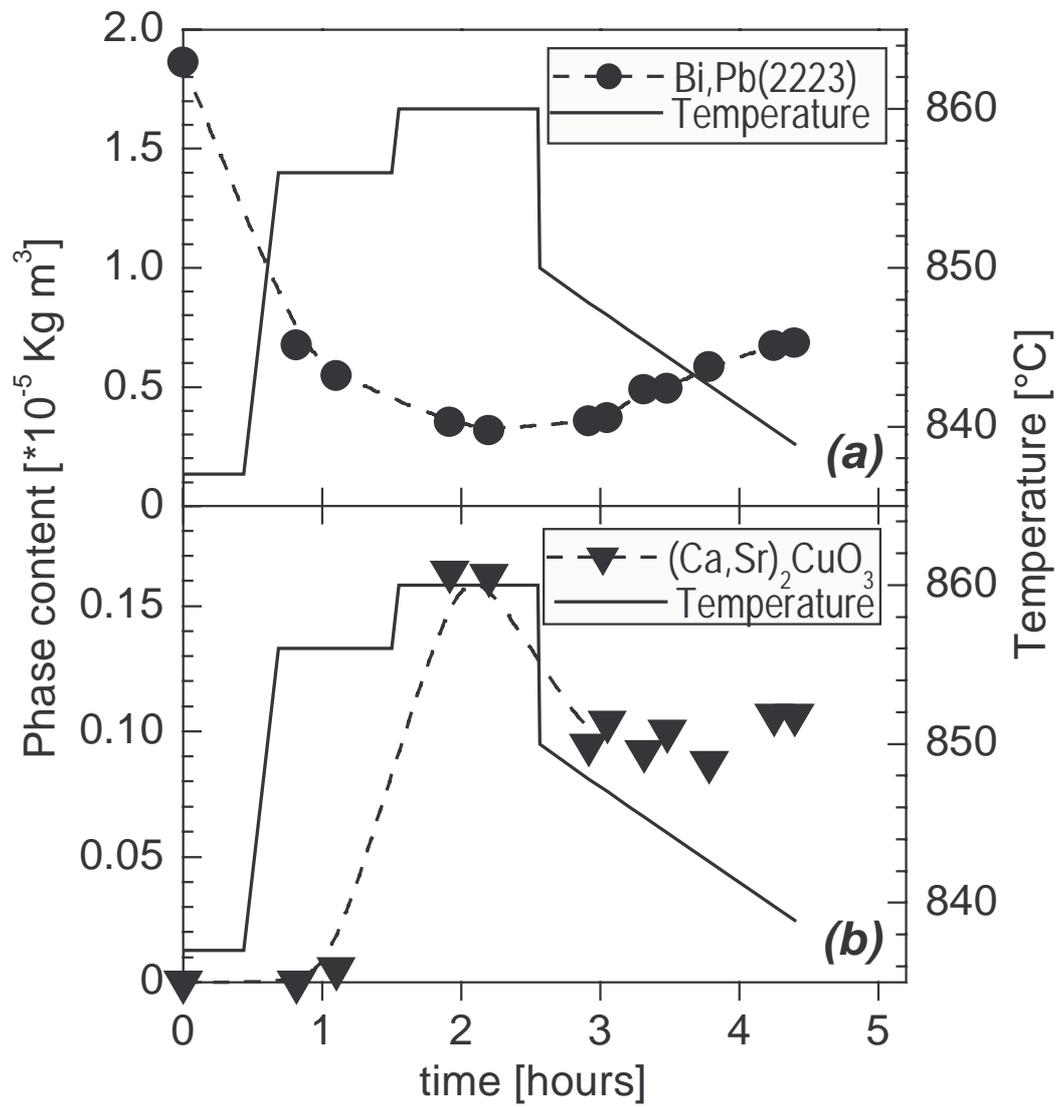

Fig. 7

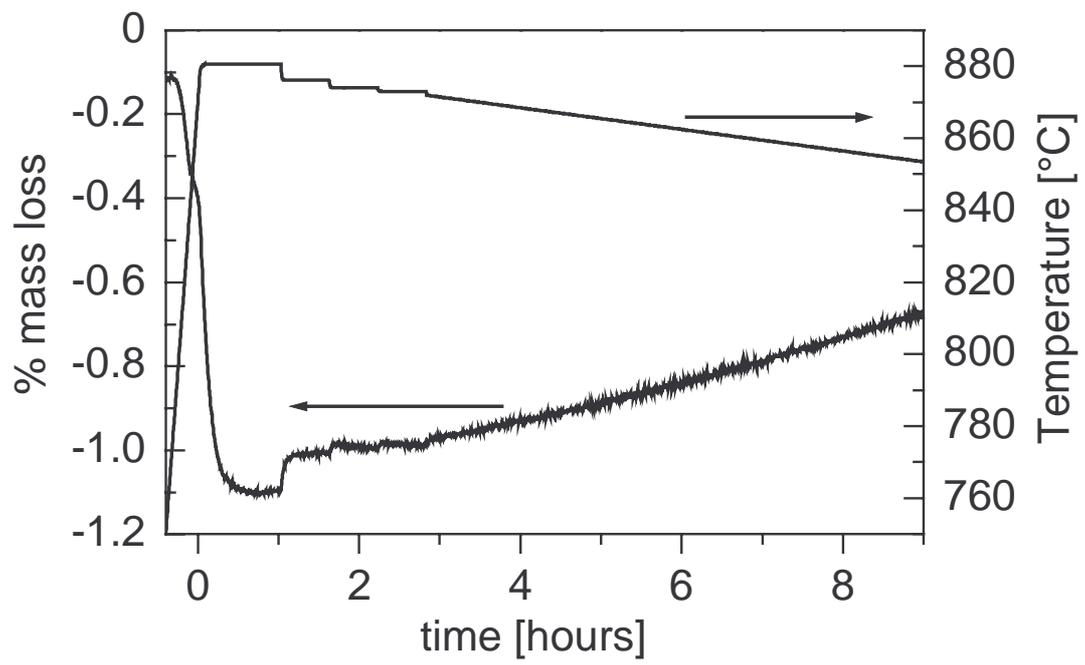

Fig. 8

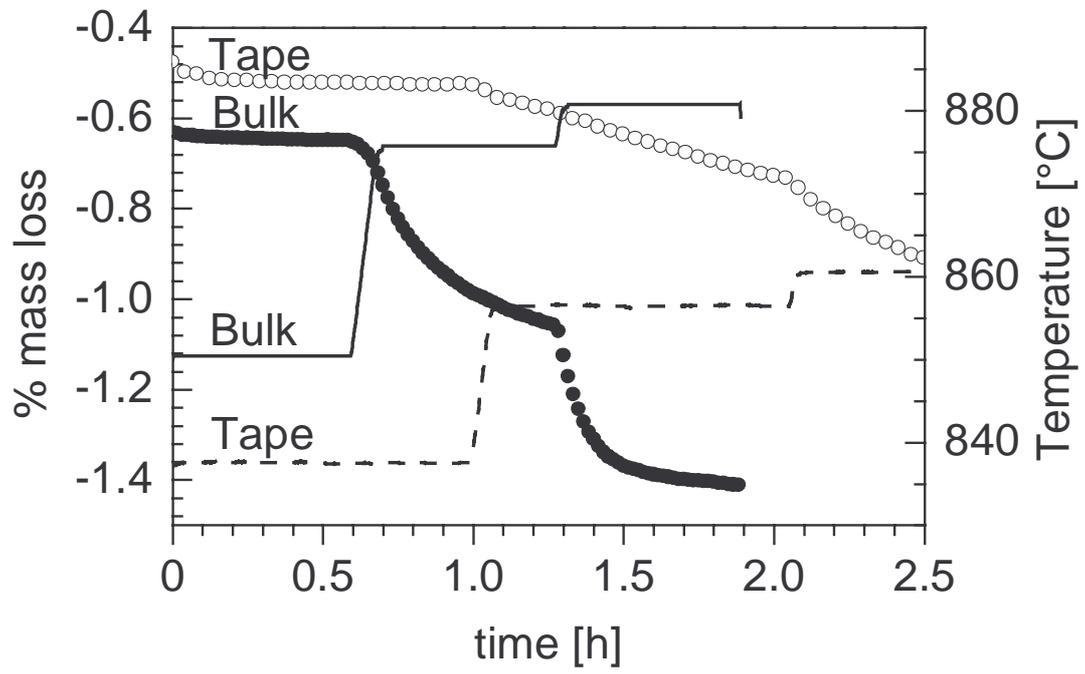

Fig. 9

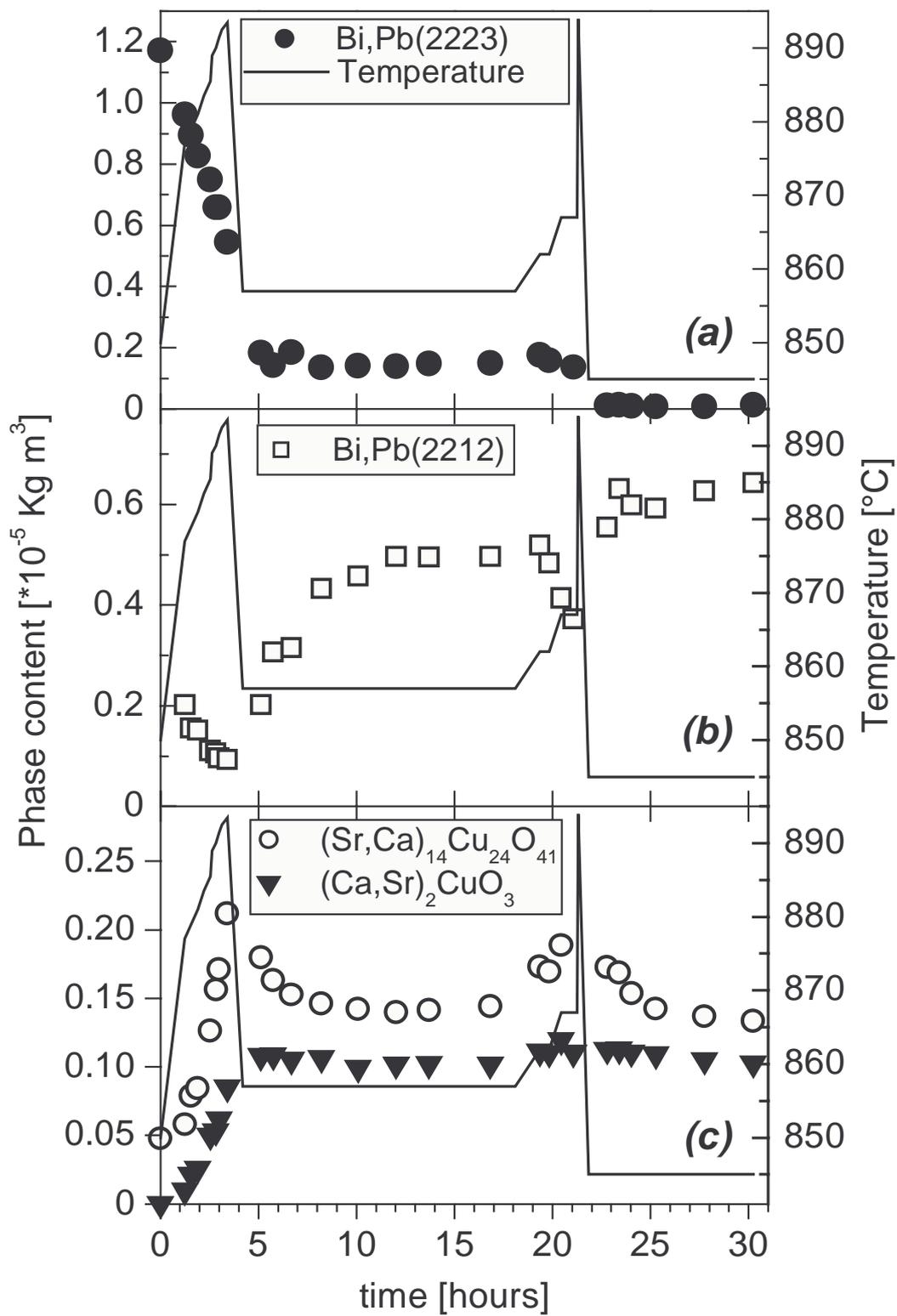

Fig. 10

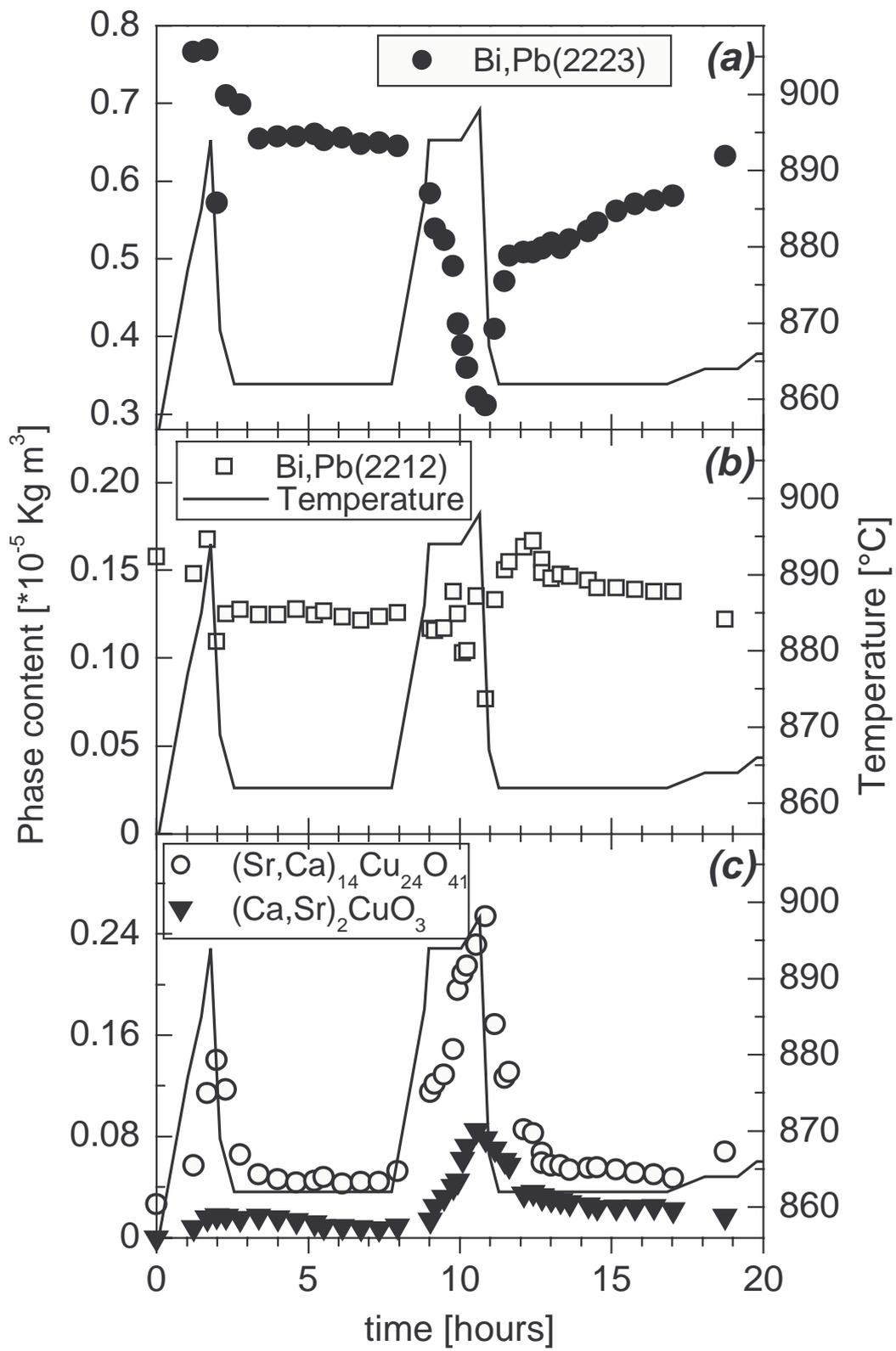

Fig. 11

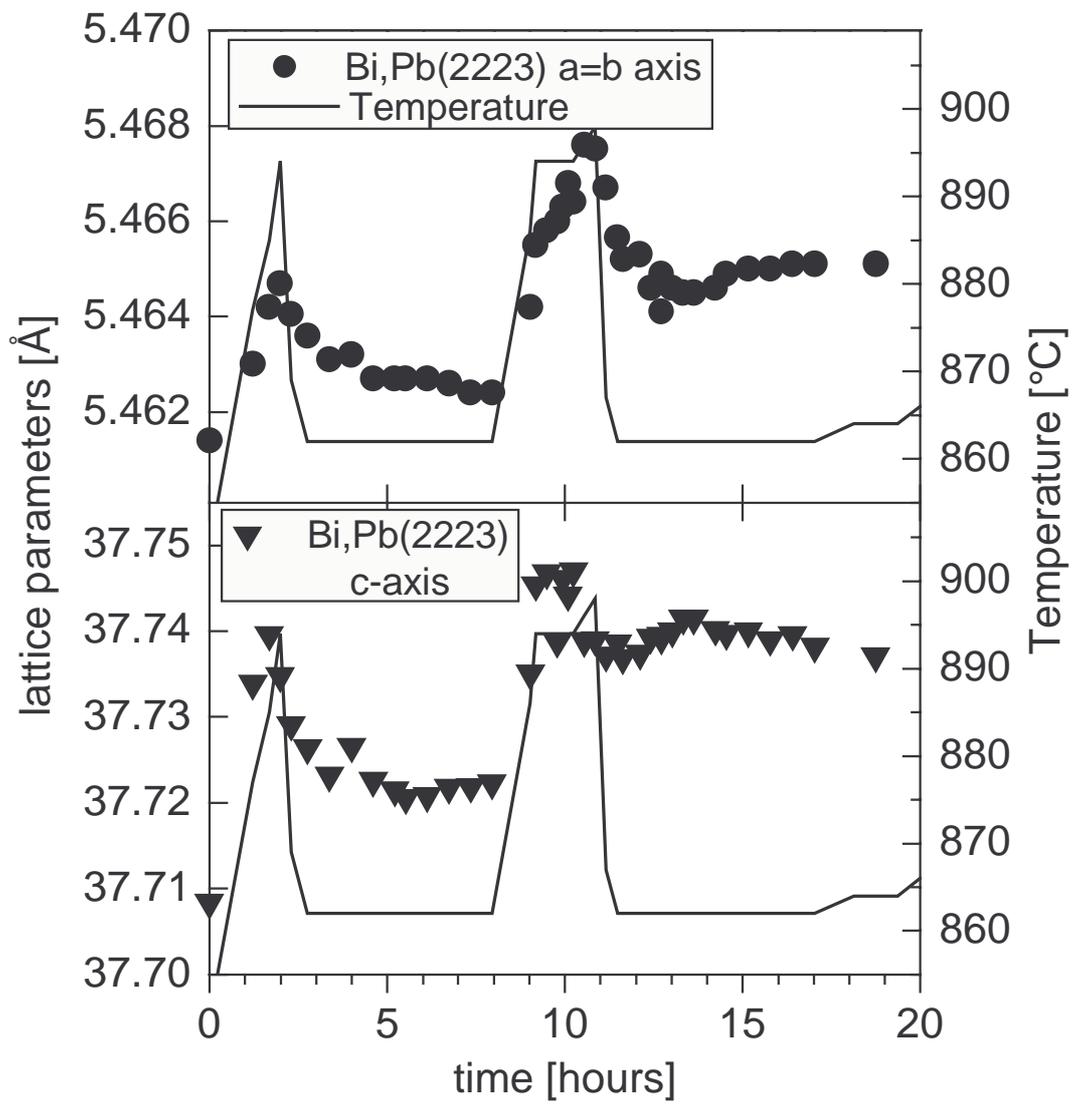

Fig. 12

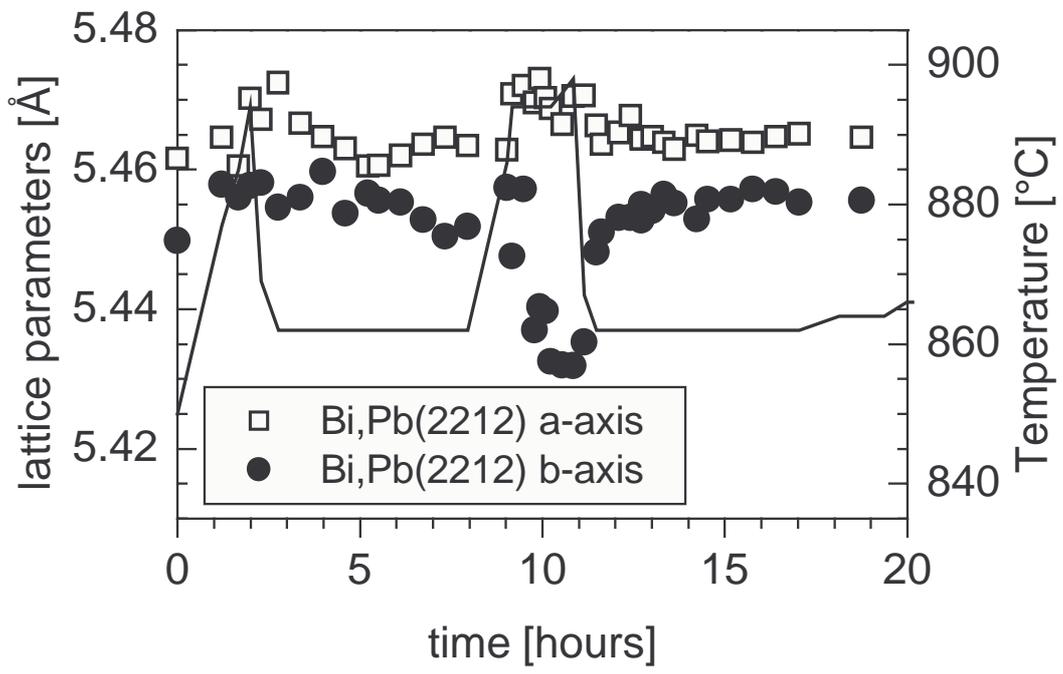

Fig. 13